\documentclass[]{spie}  %>>> use for US letter paper
\usepackage[]{graphicx}
\newcommand{\0}{{\rm {\scriptscriptstyle {0}}}}
\newcommand{\de}{{\rm {\scriptscriptstyle {DE}}}}
\title{Unknowns and unknown unknowns: \\
from dark sky to dark matter and dark energy} 

\author{Yasushi Suto
\skiplinehalf
Department of Physics, The University of Tokyo,
Tokyo 113-0033, Japan\\}

\authorinfo{Also a global scholar at Department of Astrophysical
Sciences, Princeton University, Princeton, NJ 08544, USA. ~
e-mail: suto@phys.s.u-tokyo.ac.jp}
\begin{document} 
  \maketitle 

%%%%%%%%%%%%%%%%%%%%%%%%%%%%%%%%%%%%%%%%%%%%%%%%%%%%%%%%%%%%% 
\begin{abstract}
 Answering well-known fundamental questions is usually regarded as the
major goal of science.  Discovery of other unknown and fundamental
questions is, however, even more important. Recognition that ``{\it we
didn't know anything}'' is the basic scientific driver for the next
generation. Cosmology indeed enjoys such an exciting epoch.

What is the composition of our universe ? This is one of the well-known
fundamental questions that philosophers, astronomers and physicists
have tried to answer for centuries. Around the end of the last
century, cosmologists finally recognized that ``We didn't know
anything''. Except for atoms that comprise slightly less than 5\% of
the universe, our universe is apparently dominated by unknown
components; 23\% is the known unknown (dark matter), and 72\% is the
unknown unknown (dark energy).

In the course of answering a known fundamental question, we have 
discovered an unknown, even more fundamental, question: ``What is dark
matter? What is dark energy?'' There are a variety of realistic
particle physics models for dark matter, and its experimental
detection may be within reach. On the other hand, it is fair to say
that there is no widely accepted theoretical framework to describe the
nature of dark energy. This is exactly why astronomical observations
will play a key role in unveiling its nature.

I will review our current understanding of the ``dark sky'', and then
present on-going Japanese project, SuMIRe, to discover even more
unexpected questions.
\end{abstract}

%>>>> Include a list of keywords after the abstract 


%%%%%%%%%%%%%%%%%%%%%%%%%%%%%%%%%%%%%%%%%%%%%%%%%%%%%%%%%%%%%
\section{INTRODUCTION}
\label{sec:intro} 

Darkness is the key to understanding our {\it world} in both literal and
metaphorical senses.  Philosophy, astronomy, and therefore physics
should have started by {\it thinking in the dark} in the ancient era. I
believe that this still applies now at the frontier of the contemporary
science. Serious scientific investigations for another element, another
Earth, and another life have have opened new research fields in
astronomy; dark matter, dark energy, extrasolar planet, and
astrobiology.

\begin{figure}[h]
   \begin{center}
   \begin{tabular}{c}
   \includegraphics[height=7cm]{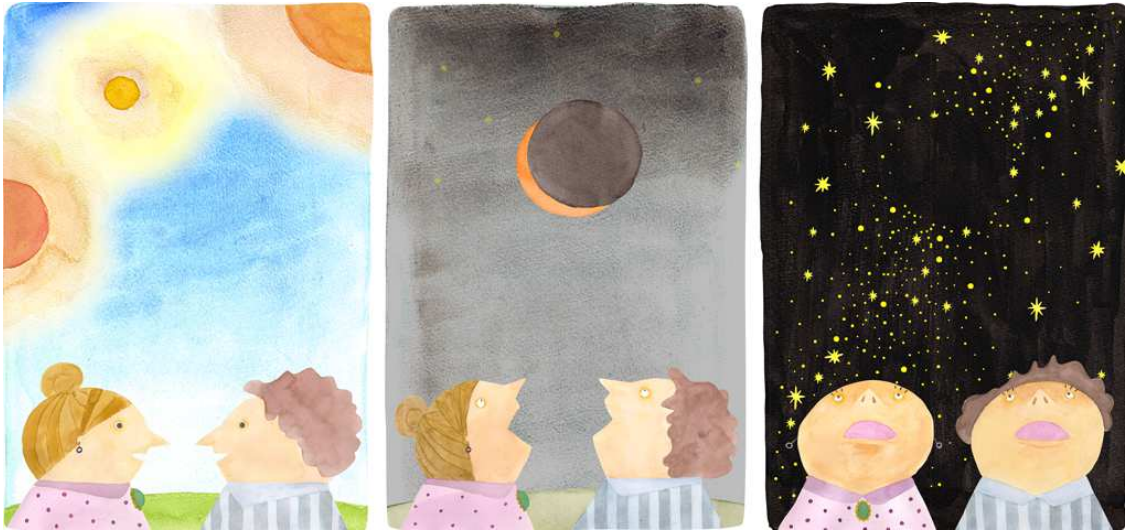}
   \end{tabular}
   \end{center}
   \caption{Nightfall: we did not know anything 
(illustration by Alisa Haba)\label{fig:nightfall}}
   \end{figure}

In his well-known novella {\sl Nightfall}\cite{nightfall}, Issac Asimov
considered a planet ``Lagash'' with six different ``Suns'' on which
``night'' comes only every 2049 years at the occasion of the total
eclipse of a Sun in the sky. People on the planet understood the real
world, for the first time, only during Lagash's brief and rare dark
night (Fig. \ref{fig:nightfall}).

\begin{quote}
\it 
``Light !'' he screamed. Aton, somewhere, was crying, whimpering
horribly like a terribly frightened child. ``Stars --  all the
Stars -- we didn't know at all. We didn't know anything.''
\end{quote}

%%%%%%%%%%%%%%%%%%%%%%%%
%%%%%%%%%%%%%%%%%%%%%%%%

Indeed the story tells us in a convincing fashion that we usually do not
realize that we do not know anything when we get so used to what we
directly see. If our planet were always covered by blue sky, we would
have never dreamed of the presence of numerous stars behind it.

It is exactly why until relatively recently we have never imagined the
fact that darkness in the night sky is dominated by something we do not
know; dark matter and dark energy.  The combined analysis of the Hubble
diagram of type Ia supernovae, the angular power spectrum of CMB (cosmic microwave background)
temperature anisotropies, and the baryon acoustic oscillation in galaxy
power spectrum among other indications unveiled the precise composition
of the present universe \cite{komatsu2010}.

Maybe even more surprisingly, baryons in the present universe have
largely evaded the direct detection so far, i.e., most of the baryons are
{\it dark}\cite{Fukugita1998}.  The majority of those {\it cosmic
dark baryons} in the present universe are thought to take the form of
the warm-hot intergalactic medium with $10^5 {\rm K}< T < 10^7 {\rm K}$
that does not exhibit any strong observational signature
\cite{Cen1999,Yoshikawa2003,Kawahara2006}.

%%%%%%%%%%%%%%%%%%%%%%%%
   \begin{figure}[h]
   \begin{center}
   \begin{tabular}{c}
   \includegraphics[height=5cm]{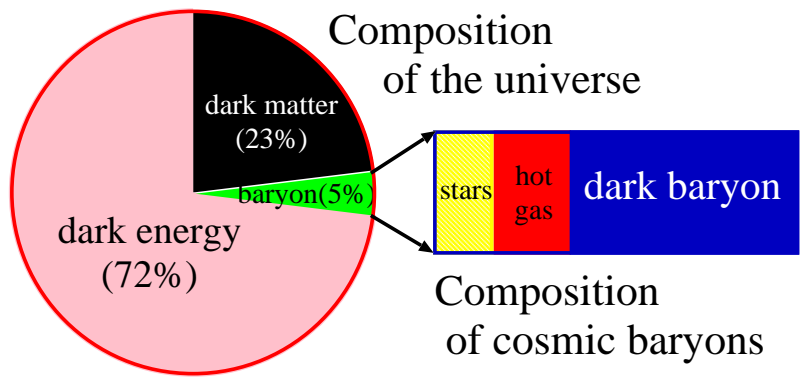}
   \end{tabular}
   \end{center}
   \caption{Known knowns, known unknowns, and unknown unknowns in the
    present universe\label{fig:piechart}}
   \end{figure}
%%%%%%%%%%%%%%%%%%%%%%%%

In summary, our universe is apparently dominated by unknown dark
components as shown in Figure \ref{fig:piechart}. The current situation
may be well described by a controversial press briefing given by former
US Defense Secretary Donald H. Rumsfeld on February 12, 2002.

\begin{quote}
\it  
There are known knowns. These are things we know that we know. There are
known unknowns. That is to say, there are things that we know we don't
know. But there are also unknown unknowns. There are things we don't
know we don't know.
\end{quote}

\begin{figure}[h]
\begin{center}
\begin{tabular}{c}
\includegraphics[height=6cm]{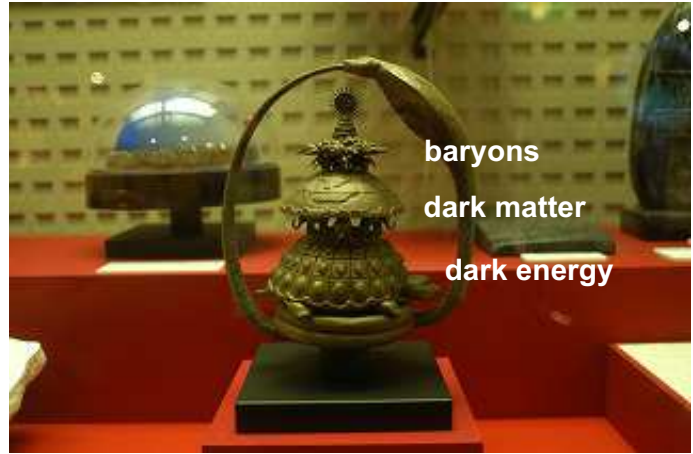}
\end{tabular}
\end{center}
\caption{Ancient Indian idea of the universe
 (courtesy of Chiharu Ishizuka, Osaka Science Museum, Japan). 
\label{fig:indian}}
\end{figure}

I am always a bit concerned by the fact that many colleagues in the
cosmology community tend to show something like Figure
\ref{fig:piechart} even proudly, indicating that our understanding of
the universe is amazingly precise. It may be true in some sense, but you
might well consider it as a paraphrase of the old Indian picture of the
universe (Fig.\ref{fig:indian}). Have we made progress at all since then
?  Maybe yes, but we have to recognize that we still do not know
something. This is why we should move on and try to unveil the nature of
the dark universe.

 In what follows, I basically focus on dark energy with particular
emphasis on our on-going project with Subaru telescope. A complementary
and more comprehensive discussion on imaging surveys of galaxies,
especially that with LSST (Large Synoptic Survey Telescope), is
presented by Prof. Tony Tyson in these proceedings.

%%%%%%%%%%%%%%%%%%%%%%%%
%%%%%%%%%%%%%%%%%%%%%%%%

%%%%%%%%%%%%%%%%%%%%%%%%%%%%%%%%%%%%%%%%%%%%%%%%%%%%%%%%%%%%%
\section{DARK ENERGY} 
\label{sec:de} 

\subsection{Cosmic Acceleration} 

The discovery of the accelerated expansion of the current universe
\cite{riess98,perlmutter99} required a fundamental modification of the
standard picture. This can be understood as follows. Let us denote the
scale factor of the universe by $a(t)$, and expand it in a Taylor series
around the present time $t_\0$:
%%%%%%%%%%%%%%%%%%%%%%%%%%%%%%%%%%%%%%%%%%%%%%%%%%%%%%%%%%%%%%%%%%%%%
\begin{eqnarray}
 a(t) = a(t_\0) + \left. \frac{d a}{dt}\right|_\0 (t-t_\0) 
+ \left. \frac{1}{2}\frac{d^2 a}{dt^2}\right|_\0 (t-t_\0)^2
 + \cdots .
\end{eqnarray}
%%%%%%%%%%%%%%%%%%%%%%%%%%%%%%%%%%%%%%%%%%%%%%%%%%%%%%%%%%%%%%%%%%%%%
The normalization of the first term $a_\0 \equiv a(t_\0)$ is arbitrary,
and thus has no direct physical meaning (conventionally it is set to
unity, $a_\0=1$). The second and third terms are related to observable
cosmological parameters, the Hubble constant $H_\0$, and the
deceleration parameter $q_\0$ as follows:
%%%%%%%%%%%%%%%%%%%%%%%%%%%%%%%%%%%%%%%%%%%%%%%%%%%%%%%%%%%%%%%%%%%%%
\begin{eqnarray}
\label{eq:h0}
H_\0 &=& \frac{1}{a_\0}\left. \frac{d a}{dt}\right|_\0 , \\
\label{eq:q0}
q_\0 &=& 
- \frac{1}{a_\0 H_\0^2}\left. \frac{d^2 a}{dt^2}\right|_\0 .
\end{eqnarray}
%%%%%%%%%%%%%%%%%%%%%%%%%%%%%%%%%%%%%%%%%%%%%%%%%%%%%%%%%%%%%%%%%%%%%
Indeed the nature of the above two parameters is intrinsically
different. The value of $H_\0$ is a purely observationally determined
parameter that cannot be predicted at all by theory. This is even true
for its signature. Observations indicate that $H_\0$ is positive,
implying that our present universe is expanding. Even if the measured
value of $H_\0$ were negative, however, it would simply mean that the
universe is contracting. So we would not be surprised even if $H_\0$
were negative; the value and signature of the expansion velocity are
just a manifestation of the initial conditions, but not of physical
laws.

This is not the case for $q_\0$. In the standard homogeneous isotropic
cosmology, one obtains
%%%%%%%%%%%%%%%%%%%%%%%%%%%%%%%%%%%%%%%%%%%%%%%%%%%%%%%%%%%%%%%%%%%%%
\begin{eqnarray}
\label{eq:d2a}
\frac{1}{a} \frac{d^2 a}{dt^2}
   =  - {4\pi G \over 3} ( \rho + 3p) ,
\end{eqnarray}
%%%%%%%%%%%%%%%%%%%%%%%%%%%%%%%%%%%%%%%%%%%%%%%%%%%%%%%%%%%%%%%%%%%%%
where $\rho=\rho(t)$ and $p=p(t)$ denote the mean density and pressure
of the universe. As long as both $\rho$ and $p$ are positive, the
left-hand-side of equation (\ref{eq:d2a}) is negative, and $q_\0$
defined by equation (\ref{eq:q0}) is positive. This is in contradiction
of the accelerated expansion of the present universe; observationally
$q_\0$ is approximately $-0.6$. The second derivative is dictated by
laws of physics, and thus it was long believed that $q_\0$ should be
positive.

One solution to the apparent contradiction is to introduce an unknown
additional constituent with $\rho_\de (>0)$ and $p_\de (<0)$. Then equation
(\ref{eq:d2a}) is rewritten as
%%%%%%%%%%%%%%%%%%%%%%%%%%%%%%%%%%%%%%%%%%%%%%%%%%%%%%%%%%%%%%%%%%%%%
\begin{eqnarray}
\label{eq:d2ade}
\frac{1}{a} \frac{d^2 a}{dt^2}
   =  - {4\pi G \over 3} (\rho + 3p + \rho_\de + 3p_\de) .
\end{eqnarray}
%%%%%%%%%%%%%%%%%%%%%%%%%%%%%%%%%%%%%%%%%%%%%%%%%%%%%%%%%%%%%%%%%%%%%
Thus if we are generous enough to accept such a strange component,
coined ``dark energy'', the cosmic acceleration can be accounted for if
$p_\de < -p - (\rho+\rho_\de)/3$.  In this context, a widely used
parameterization of the equation of state of dark energy is
%%%%%%%%%%%%%%%%%%%%%%%%%%%%%%%%%%%%%%%%%%%%%%%%%%%%%%%%%%%%%%%%%%%%%
\begin{eqnarray}
\label{eq:def-w}
w_\de \equiv p_\de / \rho_\de .
\end{eqnarray}
%%%%%%%%%%%%%%%%%%%%%%%%%%%%%%%%%%%%%%%%%%%%%%%%%%%%%%%%%%%%%%%%%%%%%
Apart from dark energy, the present universe is known to be
approximately dominated by non-relativistic matter (baryons and dark
matter) with $\rho_{\rm m}$ and negligible pressure. In this case,
equation (\ref{eq:d2ade}) reduces to
%%%%%%%%%%%%%%%%%%%%%%%%%%%%%%%%%%%%%%%%%%%%%%%%%%%%%%%%%%%%%%%%%%%%%
\begin{eqnarray}
\label{eq:d2ade2}
\frac{1}{a} \frac{d^2 a}{dt^2}
  &=&  - {4\pi G \over 3} \left[\rho_{\rm m} + (1+3w_\de) \rho_\de \right]
  \cr
&=& \frac{H_\0^2}{2} \left[\frac{\Omega_{\rm m}}{a^3} 
+ (1+3w_\de) \Omega_\de 
\exp \left(3 \int_a^{a_\0} \frac{1+w_\de}{a}da\right)
\right],
\end{eqnarray}
%%%%%%%%%%%%%%%%%%%%%%%%%%%%%%%%%%%%%%%%%%%%%%%%%%%%%%%%%%%%%%%%%%%%%
where
%%%%%%%%%%%%%%%%%%%%%%%%%%%%%%%%%%%%%%%%%%%%%%%%%%%%%%%%%%%%%%%%%%%%%
\begin{eqnarray}
\Omega_{\rm m} \equiv \frac{8\pi G}{3H_\0^2} \rho_{\rm m}(t_\0),
\quad
\Omega_\de  \equiv \frac{8\pi G}{3H_\0^2} \rho_{\de}(t_\0)
\end{eqnarray}
%%%%%%%%%%%%%%%%%%%%%%%%%%%%%%%%%%%%%%%%%%%%%%%%%%%%%%%%%%%%%%%%%%%%%
are the present ($a=a_\0$) values of the dimensionless density
parameters of matter and dark energy, respectively.

If the parameter $w_\de$ does not depend on $t$, equation (\ref{eq:d2ade2})
indicates that the deceleration parameter is written in terms of the
other parameters as
%%%%%%%%%%%%%%%%%%%%%%%%%%%%%%%%%%%%%%%%%%%%%%%%%%%%%%%%%%%%%%%%%%%%%
\begin{eqnarray}
q_\0 = \frac{\Omega_{\rm m}}{2} + \frac{1+3w_\de}{2} \Omega_\de .
\end{eqnarray}
%%%%%%%%%%%%%%%%%%%%%%%%%%%%%%%%%%%%%%%%%%%%%%%%%%%%%%%%%%%%%%%%%%%%%
All the currently available data are surprisingly consistent with
$w_\de=-1$, which corresponds to the cosmological constant, $\Lambda$,
that Einstein originally introduced in 1917 for a completely different
reason. In this case, the condition for the cosmic acceleration
($q_\0<0$) is
%%%%%%%%%%%%%%%%%%%%%%%%%%%%%%%%%%%%%%%%%%%%%%%%%%%%%%%%%%%%%%%%%%%%%
\begin{eqnarray}
\Omega_\de=\Omega_\Lambda > \frac{\Omega_{\rm m}}{2} .
\end{eqnarray}
%%%%%%%%%%%%%%%%%%%%%%%%%%%%%%%%%%%%%%%%%%%%%%%%%%%%%%%%%%%%%%%%%%%%%
Therefore the cosmic acceleration implies that our present universe
should be dominated by a peculiar unknown component that behaves very
similarly as Einstein's cosmological constant; various observations
point to $\Omega_\Lambda\approx 0.72$ and $\Omega_{\rm m} \approx
0.28$. Is dark energy really a cosmological constant ? In other words,
{\it $w_\de =-1$ or not, that is the question}.  This is one of the
crucial questions on dark energy that we would like to answer within a
decade from now.

Another solution is to assume that general relativity is not strictly
valid on cosmological scales, commonly referred to as modified gravity
theories. Equation (\ref{eq:d2a}) with $p=0$ may be rewritten as
%%%%%%%%%%%%%%%%%%%%%%%%%%%%%%%%%%%%%%%%%%%%%%%%%%%%%%%%%%%%%%%%%%%%%
\begin{eqnarray}
\label{eq:d2a-inv2}
\frac{d^2 a}{dt^2}
   =  - \frac{G}{a^2} \left( \frac{4\pi\rho a^3}{3} \right)
   =  - \frac{GM(<a)}{a^2} ,
\end{eqnarray}
%%%%%%%%%%%%%%%%%%%%%%%%%%%%%%%%%%%%%%%%%%%%%%%%%%%%%%%%%%%%%%%%%%%%%
indicating that this is nothing but the inverse square law of
gravitation. While dark energy models abandons the idea that our
universe is dominated by normal components with positive pressure,
modified gravity models abandon the validity of the inverse square law
on cosmological scales. 

The two different solutions to the cosmic acceleration should lead to
different predictions on growth history of structures in the universe,
which can be tested against the galaxy survey data as will be discussed
below.

\subsection{Baryon Acoustic Oscillation (BAO)} 

The baryon acoustic oscillation (BAO) is an oscillation of photon-baryon
fluid imprinted in the matter spectrum, which was detected in the SDSS
and 2dFGRS galaxy distribution in 2005 (e.g.,
Refs.~\citenum{Eisenstein2005,Cole2005}).  The characteristic scale
imprinted in the BAO is basically the sound horizon at recombination:
%%%%%%%%%%%%%%%%%%%%%%%%%%%%%%%%%%%%%%%%%%%%%%%%%%%%%%%%%%%%%%%%%%%%%%%% 
\begin{eqnarray}
r_s(z_{\rm rec}) &=& \int_{z_{\rm rec}}^{\infty}\,\frac{dz\,c_s(z)}{H(z)} 
= \frac{2}{3k_{\rm eq}}\sqrt{\frac{6}{R_{\rm eq}}}\ln
\frac{\sqrt{1+R_{\rm rec}}+\sqrt{R_{\rm rec}+R_{\rm eq}}}{1+\sqrt{R_{\rm
eq}}}
\cr
&\approx&
147(\Omega_mh^2/0.13)^{-0.25}(\Omega_bh^2/0.024)^{-0.08}{\rm Mpc},
\label{eq:soundhorizon}
\end{eqnarray}
%%%%%%%%%%%%%%%%%%%%%%%%%%%%%%%%%%%%%%%%%%%%%%%%%%%%%%%%%%%%%%%%%%%%%%%% 
where $c_s(z)$ is the sound speed at redshift $z$, and $z_{\rm rec}$ is
the redshift at recombination ($\simeq1089$). In the second equality,
$k_{\rm eq}$ is the horizon scale at the matter-radiation equality
epoch, $z_{\rm eq}$, $R_{\rm rec}=R(z_{\rm rec})$ and $R_{\rm eq}
=R(z_{\rm eq})$ are the ratio of the baryon to photon momentum densities
at $z_{\rm rec}$ and $z_{\rm eq}$. Finally the last equality is an
approximate fit where $\Omega_m$ and $\Omega_b$ are the density
parameters of matter and baryon, and $h$ is the  Hubble constant
in units of 100km$\,$s$^{-1}$Mpc$^{-1}$.

\begin{figure}[h]
\begin{center}
\includegraphics[width=6cm]{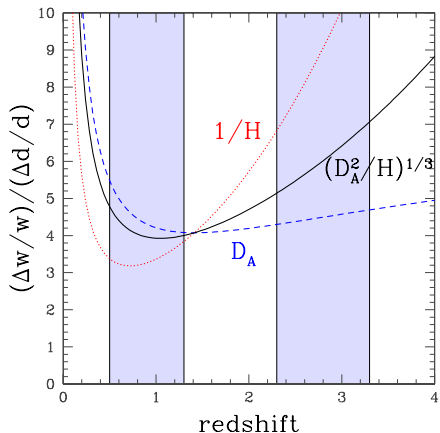}
\end{center}
\caption{The error propagation from measured scales, $d$, to the dark 
energy equation of state parameter, $w_{\rm DE}$, as a function of 
redshift. We choose $1/H(z)$ (dotted), and $D_A(z)$ (dashed) for $d$,
which correspond to the separations parallel and 
perpendicular to the line-of-sight direction. We also plot
the three dimensional average, $(D_A^2(z)/H(z))^{1/3}$ (solid) for $d$. 
The shaded regions indicate the targeted redshift 
ranges of a future galaxy survey, WFMOS (reproduced 
from Ref.~\citenum{nishimichi07}).}
\label{fig:dwdr}
\end{figure}

The BAO length scale is a good tracer of $w_\de$, and also of modified
gravity theories\cite{Yamamoto2006}.  The BAO scale measured
from a redshift survey of galaxies provides estimates of the angular
diameter distance, $D_A(z)$, and the inverse of the Hubble parameter,
$1/H(z)$, which correspond to the scales perpendicular and parallel to
the line-of-sight direction, respectively. They in turn can be
translated into the estimate of $w_{\rm DE}$. Figure \ref{fig:dwdr}
shows how the fractional errors of three important scales, the angular
diameter distance $D_{A}(z)$, the inverse of Hubble parameter $1/H(z)$
and their average over three dimensions, $(D_A^2(z)/H(z))^{1/3}$,
propagate to that of $w_{\rm DE}$.  The two shaded regions show the
approximate targeted redshift ranges of a galaxy redshift survey, WFMOS
(Wide-field Fiber-fed Multi-Object Spectrograph), which is basically
the same as SuMIRe PFS redshift surveys discussed in \S \ref{sec:sumire}.
Typically a ratio of $\Delta w/w$ and $\Delta d/d$ around $z=1$ ranges
from 3 to 5.

%%%%%%%%%%%%%%%%%%%%%%%%%%%%%%%%%%%%%%%%%%%%%%%%%%%%%%%%%%%%%%%%%%%%%%%% 
%%%%%%%%%%%%%%%%%%%%%%%%%%%%%%%%%%%%%%%%%%%%%%%%%%%%%%%%%%%%%%%%%%%%%%%% 

Figure \ref{fig:dwdr} indicates that determining $w_{\rm DE}$ within
$3$\% accuracy requires the sub-percent accuracy/precision in
determining the BAO scale. This is a very demanding requirement for
on-going galaxy surveys, and the reliable theoretical predictions to
that level of accuracy are non-trivial to make. This is why
various authors have been developing different semi-analytic models to
compute power spectra of galaxies taking account of nonlinear
gravitational evolution, biasing with respect to dark matter, and
redshift-space distortion.  Since the standard perturbation
theory(PT)\cite{suto91,makino92,Jeong2006,Jeong2009} breaks down even at
a relatively weak nonlinear regime, those approaches are exploring
different new re-summation techniques so that a class of infinite series
of higher-order corrections can be incorporated even empirically by
modifying the standard PT\cite{crocce09,matsubara09}.

\begin{figure}[t]
\begin{center}
\includegraphics[height=8.5cm]{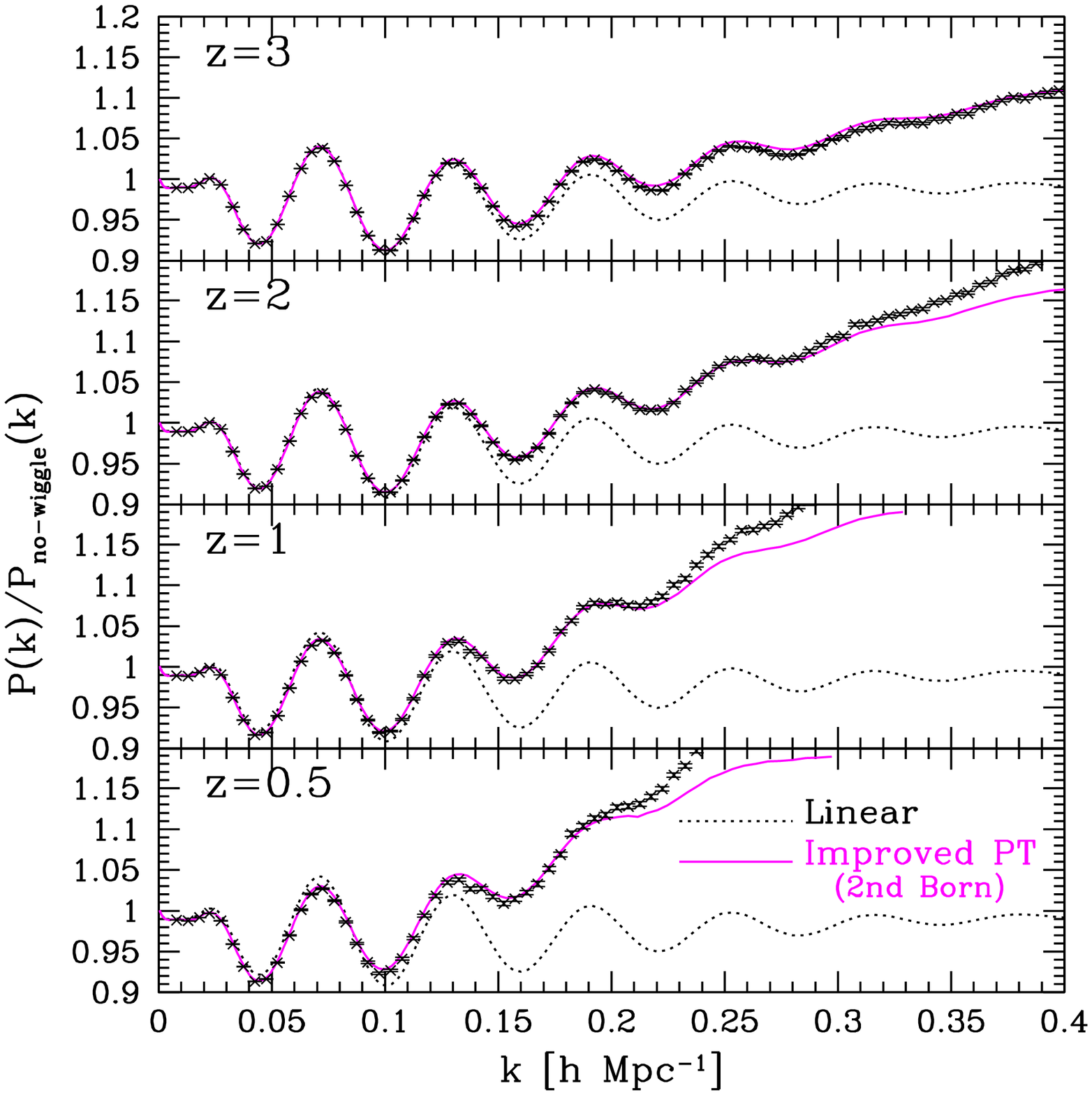}
\includegraphics[height=8.5cm]{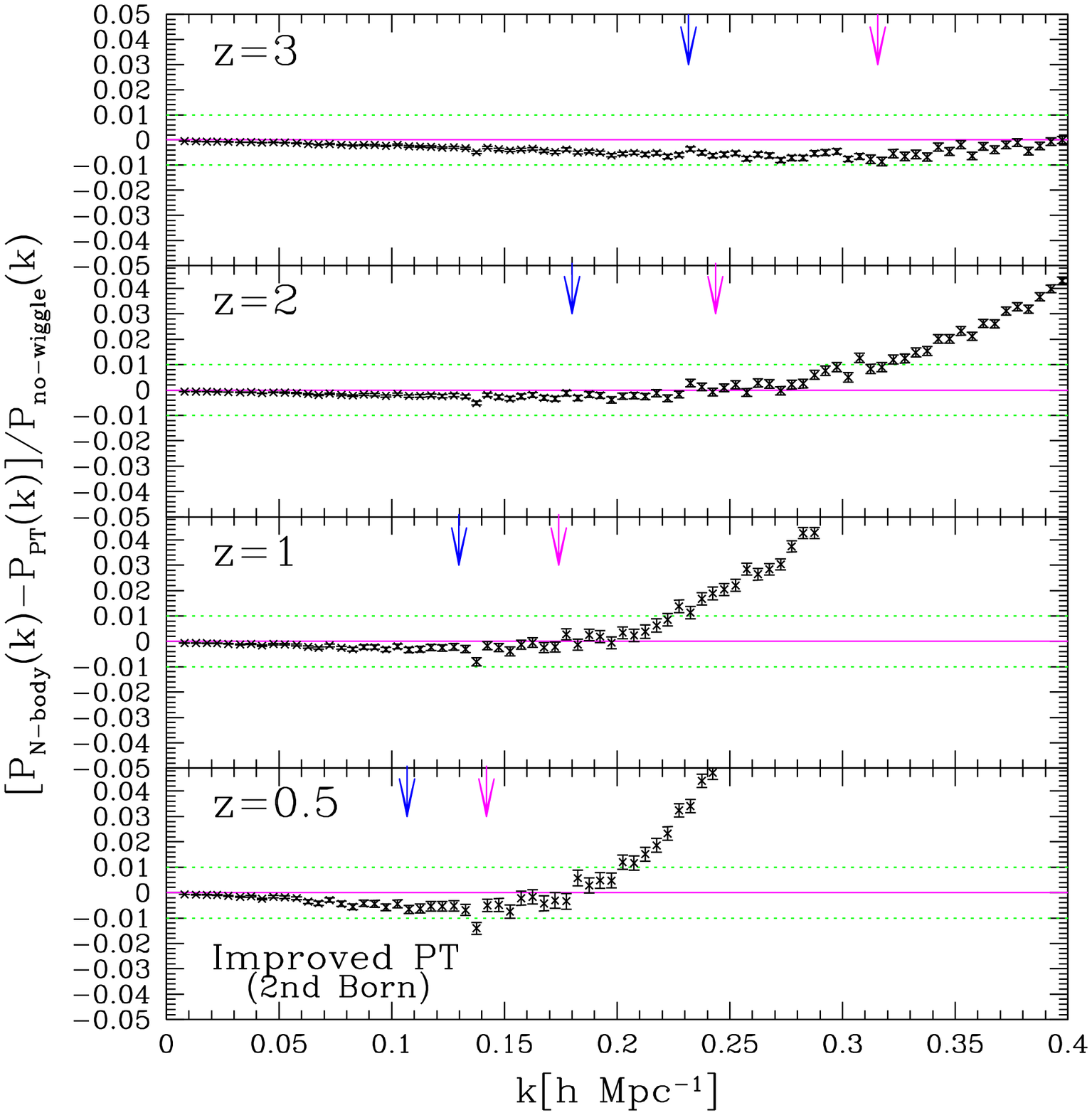}
\end{center}
\caption{Comparison between N-body results and improved PT predictions.
From top to bottom, the results at $z=3$, $2$, $1$ and $0.5$ are shown.
The improved PT predictions plotted here include the corrections up to
the second-order Born approximation of the mode-coupling term, $P^{\rm
MC2}$.  {\it Left}: ratio of power spectrum to the smoothed reference
spectra, $P(k)/P_{\rm no\mbox{-}wiggle}(k)$. Solid and dotted lines are
improved PT and linear theory predictions, respectively.  {\it Right}:
difference between N-body and improved PT results normalized by the
no-wiggle formula, $[P_{\rm N\mbox{-}body}(k)-P_{\rm PT}(k)]/P_{\rm
no\mbox{-}wiggle}(k)$.  In each panel, vertical arrows represent the
wavenumber $k_{1\%}$ for the leading-order predictions of standard and
improved PT from left to right (reproduced from
Ref.~\citenum{taruya2009}).  \label{fig:taruyafig7}}
\end{figure}

One of the most successful methods (see
Refs.~\citenum{carlson09,nishimichi2009} for systematic comparison
between simulations and different analytic methods) is based on the
closure theory widely used in describing turbulence. Atsushi Taruya and
his collaborators applied the methodology to cosmological perturbation
theory for the first time\cite{taruya2008,hiramatsu09}. Their results of
dark matter in real space\cite{taruya2009} are shown in
Figure~\ref{fig:taruyafig7} as specific examples of the BAO features.
The comparison between the model and simulations indicate that the
sub-percent level accuracy is now achieved for dark matter spectra in
real space. Nevertheless the clustering statistics measured from actual
galaxy surveys have to be computed in redshift space with galaxy
biasing.  It is also important to formulate those methods in modified
gravity theories so as to see any departure from dark energy
models\cite{koyama09}.  Thus there are still many aspects that should be
improved theoretically, and the efforts towards the direction are being
explored intensively.

%%%%%%%%%%%%%%%%%%%%%%%%%%%%%%%%%%%%%%%%%%%%%%%%%%%%%%%%%%%
%%%%%%%%%%%%%%%%%%%%%%%%%%%%%%%%%%%%%%%%%%%%%%%%%%%%%%%%%%%

%%%%%%%%%%%%%%%%%%%%%%%%%%%%%%%%%%%%%%%%%%%%%%%%%%%%%%%%%%%%%
\section{Subaru Measurement of Images and Redshifts (SuMIRe)} 
\label{sec:sumire}

The Subaru Measurement of Images and Redshifts (SuMIRe) of the universe
is an international galaxy survey project, which comprises a wide-field
imaging survey with Hyper Suprime-Cam (HSC) and a redshift survey with
Prime-Focus Spectrograph (PFS).  One of the major goals of SuMIRe is to
probe the evolution of cosmic structures, and thereby to unveil the
nature of dark energy and/or test general relativity at cosmological
scales. For that purpose, we combine four different methods; weak
gravitational lensing, BAO, cluster counting statistics, and the
redshift -- magnitude relation of distant type-Ia supernovae.  The
imaging survey with HSC will observe a billion of galaxies, and the
spectroscopic survey with PFS will determine redshifts of a few millions
of galaxies (and QSOs as well). The combined SuMIRe survey will complete
a three-dimensional map of dark matter, galaxies and clusters over a
wide range of redshifts up to $z \approx 1.6$.

\subsection{Brief history of SuMIRe}

The HSC was originally proposed by Satoshi Miyazaki and his colleagues
at NAOJ (National Astronomical Observatory of Japan) as one of next-
generation instruments on Subaru 8.2 m telescope at Mauna Kea,
Hawaii. They completed the conceptual design and concluded that the
field size of up to 2 degree in diameter is technically
feasible~\cite{2006SPIE.6269E...9M}. Based on their studies, the first
major funding proposal (P.I. Hiroshi Karoji, NAOJ) was submitted in 2005
to Japanese Ministry of Education, and was approved as a 6 year project
(September 2006 -- March 2012). Later ASIAA (Academia Sinica Institute
of Astronomy and Astrophysics) and Department of Astrophysical Sciences,
Princeton University officially joined the project.  We expect to have
the HSC first light at the end of 2011, and hope to start imaging
surveys around mid-2012.

In contrast, PFS has a long and complicated history. The original idea
of PFS should be traced back to KAOS (Kilo-Aperture Optical
Spectrograph), a prime focus wide-field multi-object fiber spectrograph
concept for one of the Gemini Telescopes in 2002. In 2003, KAOS was
renamed as WFMOS (Wide-field Fiber-fed Multi-Object Spectrograph), and
proposed as one of Gemini's next-generation instruments at Aspen.  Soon
later, however, it was recognized that the Gemini telescope cannot
mechanically support such a heavy instrument at Prime Focus. In 2004
Hiroshi Karoji, the director of Subaru Observatory at that time,
suggested a possibility that WFMOS should be installed on Subaru.
Karoji's suggestion was taken very seriously by a group of people in the
Gemini community, and White paper by the WFMOS Feasibility Study Dark
Energy Team\cite{glazebrook05} was released in 2005.
 
In December 2005, the Gemini observatory started a WFMOS conceptual
design study. While it was once suspended in May 2006 due to the budget
uncertainty, it was resumed in May 2007, and in February 2009, a
JPL/Caltech team (P.I. Richard Ellis, Caltech) was selected for the
design group.  In May 2009, however, the Gemini Board finally decided to
cancel WFMOS due to the budget problem.

Just after that, the Japanese government announced the stimulus funding
program that will select 30 top-researchers in Japan and provide
$2.7\times 10^{11}$ yen in total for the next 5 years.  The director of
IPMU (Institute of Physics and Mathematics of the Universe, The
University of Tokyo), Hitoshi Murayama, proposed ``Revealing the past
and future of the universe: unveiling the nature of dark matter and dark
energy with ultra-wide-field imaging and spectroscopy'', which was
approved on September 5, 2009, 6 days after the historical defeat of
Japan Liberal Democratic Party  at the Japanese general election on
August 30, 2009.

The proposal is a joint project of HSC and PFS, and indeed is the
current SuMIRe project. Incidentally ``Sumire'' is the Japanese name of
a flower (``violet'' in English). I invented the acronym first, and
later David Spergel of Princeton University came up with the full name
``SUbaru Measurement of Images and REdshift of the universe'' for it.

Of course, this is not the end of the story.  On September 16, 2009, new
Japanese government started, and on October 17, 2009, it reduced the
total budget of the stimulus package (for 30 projects) from $2.7\times
10^{11}$ yen to $1.0 \times 10^{11}$ yen.  Nevertheless we are supposed
to complete PFS in the international collaboration scheme in any way.
At the time of this writing, therefore, we are pursuing a variety of
possibilities to secure the funding necessary for PFS, and hope to start
the collaboration within a year or so.

\subsection{Hyper Suprime-Cam (HSC)} 

\begin{figure}[h]
\begin{center}
   \begin{tabular}{c}
   \includegraphics[height=8cm]{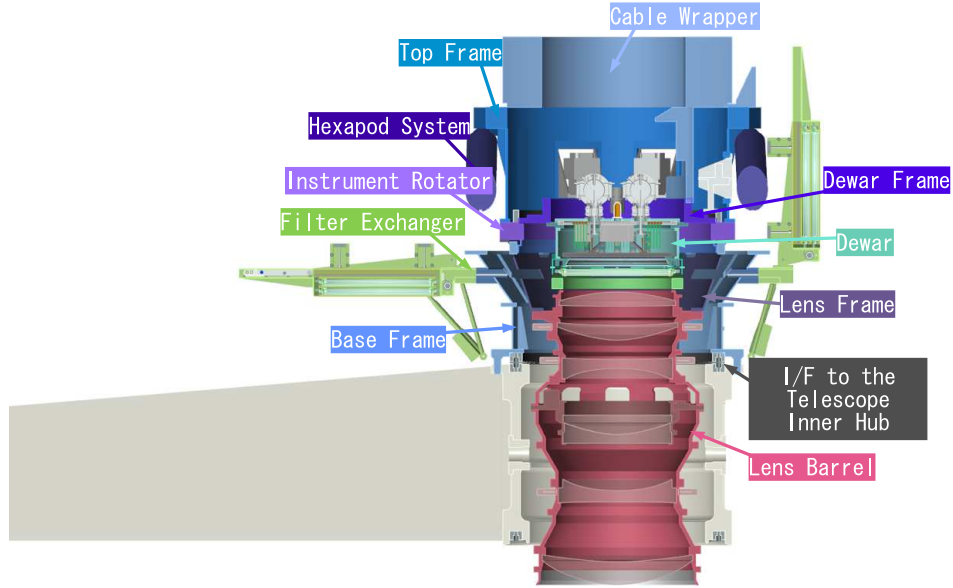}
   \end{tabular}
\end{center}
\caption{Cross sectional view of Hyper Suprime-Cam unit mounted on
  telescope's top end hub (reproduced from Ref.~\citenum{miyazaki09}).}
\label{fig:hsc}
\end{figure}

{\it Hyper Suprime-Cam} (HSC) is an upgrade of the existing
camera {\it Suprime-Cam} (SC) installed at the Prime Focus of Subaru
Telescope. The field-of-view of HSC is nearly an order-of-magnitude
larger than than of SC, while maintaining the equivalent image quality.
Combined with the fact that Subaru's median seeing is better than 0.7
arcsec (FWHM), HSC will be a very unique and powerful instrument
particularly for weak lensing survey.  The basic specifications of HSC
are summarized in Table~\ref{tab:hsc}, and the configuration of its
major components are shown in Figure \ref{fig:hsc} (courtesy of Satoshi
Miyazaki). See Ref.~\citenum{miyazaki09} for further details of HSC.

The HSC wide-field, multi-color ($grizy$) survey covering $1500\sim
2000$ square degree sky is expected to start in mid-2012 for 5 years.
The depths of the survey correspond to $g<26$, $r<25.9$, $i<25.8$,
$z<25$, and $y<24$ ($5\sigma$ limit), respectively, and probe galaxies
up to $z=1.5$.  The planned dark energy experiments with the HSC surveys
(and also with PFS) include weak lensing, cluster counting statistics,
and SNe, which are summarized in Table~\ref{tab:de}.

%%%%%%%%%%%%%%%%%%%%%%%%%%%%%%%%%%%%%%%%%%%%%%%%%%%%%%%%%%%%%%%%%
\begin{table}[h]
\begin{center}
\small
\begin{tabular}{|l|l|l|} \hline
Field of View & 1.5 degree diameter 
& Vignetting allowed up to 25 \% at the edge \\ 
& & Dead area (CCD gap) fraction $\leq$ 5 \% \\ \hline
Instrument PSF size &  $\leq$ 0''.4  for g', r', i', z'& FWHM
,
Telescope Elevation $>$ 30 deg.\\ \hline
Pixel scale & $\leq$ 0''.2 /pix & \\ \hline
System Throughput & $\geq$ 50 \% for g'
& Primary Mirror $\times$ Wide Field Corrector $\times$Filter$\times$CCD\\ 
& $\geq$ 65 \% for r' & at the center of the field\\
& $\geq$ 65 \% for i' & \\
& $\geq$ 40 \% for z' & \\
& $\geq$ 20 \% for Y  & \\ \hline
Minimum Exposure time & 2 sec (1 sec goal) 
& Time accuracy $\leq$ $\pm$ 1 \% \\ \hline
Min. interval of Exposures & 20 sec (15 sec goal) 
& Including CCD readout and wipe \\
& & pointing change \\ \hline
Min. number of filter holders & 4 & \\
Filter Exchange Time & $<$  10 minutes & \\
\hline
\end{tabular}
\end{center}
\caption{Specifications of Hyper Suprime-Cam 
(reproduced from Ref.~\citenum{miyazaki09}).}
\label{tab:hsc}
\vspace*{0.5cm}
\end{table}
%%%%%%%%%%%%%%%%%%%%%%%%%%%%%%%%%%%%%%%%%%%%%%%%%%%%%%%%%%%%%%%%

%%%%%%%%%%%%%%%%%%%%%%%%%%%%%%%%%%%%%%%%%%%%%%%%%%%%%%%%%%%%%%%%
%%%%%%%%%%%%%%%%%%%%%%%%%%%%%%%%%%%%%%%%%%%%%%%%%%%%%%%%%%%%%%%%

%%%%%%%%%%%%%%%%%%%%%%%%%%%%%%%%%%%%%%%%%%%%
\begin{table}[h]
\begin{center}
\begin{tabular}{|l|l|l|}
\hline
Methods & Dominant systematic errors & SuMIRe approaches\\ \hline
Weak lensing & Photo-$z$ errors & PFS data (spectroscopic calibration
	 sample)\\
&Shape measurements & Subaru image-quality,
 methods \& algorithms\\ 
& Small-scale clustering &A suite of high-resolution simulations \\
\hline
Clusters& Mass-observable relations& Sunyaev-Zel'dovich-optical-lensing cross
	 calibration\\
& Selection function & PFS spectroscopic calibration of BCGs\\ \hline
BAO & Galaxy biases & Lensing-galaxy cross-correlations \\
& Nonlinearities & A suite of simulations \& refined analytical
	 methods\\ \hline 
Type-Ia supernovae & Photometric calibration & Calibration strategy,
spectroscopic calibration\\ \hline
\end{tabular}
\end{center}
\caption{SuMIRe Dark Energy Experiments (courtesy of Masahiro Takada and
 John D. Silverman)}
\label{tab:de}
\end{table}
%%%%%%%%%%%%%%%%%%%%%%%%%%%%%%%%%%%%%%%%%%%%
\clearpage

\subsection{Prime Focus Spectrograph (PFS)} 

The basic design of PFS is as follows; in total $2000$--$3000$ fibers to
target 2200 galaxies per field-of-view. The spectral resolution is
$R\simeq 3000$ for wavelength 600--1000nm in order to survey galaxies
over redshift range $0.6< z < 1.6$ and to detect absorption/emission
lines of each galaxies with enough signal-to-noise ratios ($S/N >
5$--$10$).  Thus it is possible to perform a multi-object spectroscopic
survey of early-type galaxies and emission line galaxies up to $z\approx
1.6$.

The PFS will carry out another dark energy experiment through BAO from
the measured three-dimensional galaxy distribution, which bears a great
synergy with the HSC survey.  The deep, multi-color HSC data sets can
deliver an ideal catalog to find target galaxies for the PFS redshift
survey based on the same telescope.  To maximize the synergy of HSC and
PFS surveys, we choose, as the target galaxies, red galaxies (BCGs and
early-type galaxies) in cluster regions and star-forming blue
galaxies. By targeting emission-line galaxies exhibiting [OII] (3726\AA,
3729\AA) with red-band sensitive CCD chips, we can probe the
three-dimensional galaxy distribution over a wide range of redshifts,
$0.6 < z < 1.6$.  The spectral resolution of $R\approx 3000$ is
important to discriminate [OII] emission lines from OH sky lines in red
bands beyond 8000\AA. The resulting survey is complementary to the
ongoing BOSS (Baryon Oscillation Spectroscopic Survey) with redshift
range $0.4 < z < 0.65$ and 10000 square degree area (mostly in the
northern hemisphere sky); we can combine the spectroscopic data sets
from the BOSS and PFS surveys to accurately estimate photometric
redshift errors and cluster selection function. This will significantly
reduce the systematic errors in the weak lensing and cluster
experiments.

\begin{figure}[h]
\begin{center}
   \includegraphics[height=8cm]{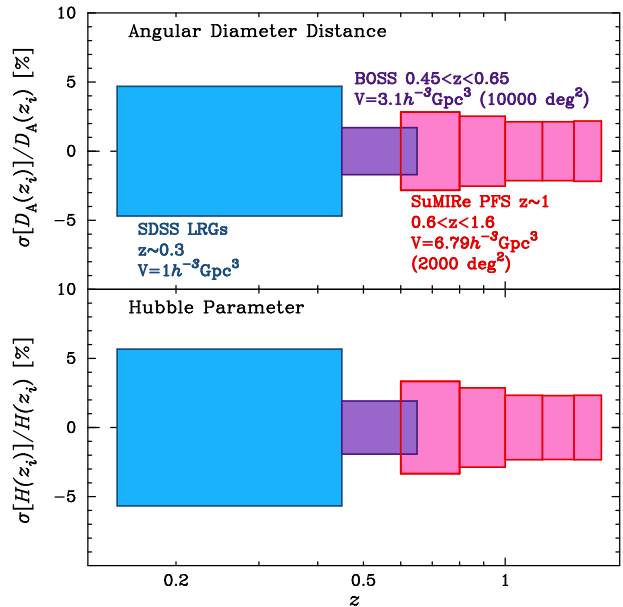} 
\end{center}
\caption{Fractional
errors in measuring the angular diameter distance and the Hubble
expansion rate for each redshift slices for the different BAO surveys,
SDSS, BOSS and PFS (courtesy of Masahiro Takada).}
\label{fig:da-h}
\end{figure}

Figure~\ref{fig:da-h} shows the expected accuracies of measuring the
angular diameter distances and the Hubble expansion rate for each
redshift slices. The PFS BAO survey can achieve a few percent accuracies
of the distance measurements and is quite complementary to the SDSS and
BOSS surveys, yielding a wider coverage of redshifts out to $z\simeq
1.6$. The dark energy contribution to the cosmic expansion is thought to
be insignificant at $z > 1$, if dark energy is the cosmological
constant, and therefore the PFS survey combined with the SDSS and BOSS
can open up a window to test early dark energy models up to $z\simeq
1.6$.

Thus the SuMIRe (HSC + PFS) survey will aim at probing the nature of
dark energy and/or testing gravity theories via weak lensing, BAO,
galaxy clustering statistics, galaxy clusters, and supernovae, in
addition to at constraining or even detecting neutrino mass and
primordial non-Gaussianity for the first time.

%%%%%%%%%%%%%%%%%%%%%%%%%%%%%%%%%%%%%%%%%%%%%%%%%
%%%%%%%%%%%%%%%%%%%%%%%%%%%%%%%%%%%%%%%%%%%%%%%%%

%%%%%%%%%%%%%%%%%%%%%%%%%%%%%%%%%%%%%%%%%%%%%%%%%%%%%%%%%%%%%
\section{CONCLUSION} 

Observational cosmology in the last decade clearly reminded us of the
fact that one could have never imagined what dominates our world without
dark nights. This may sound quite obvious now, but we do not seem to
have fully appreciated the fact until very recently.  The discovery of
the cosmic acceleration indicates either an unknown component in the
universe or an unknown law of gravity on cosmological scales. This poses
a previously unknown and fundamental question in our world. Various
on-going and up-coming international galaxy survey projects will answer
the question in the next decade by taking photos of darkness, exactly in
the same spirit of Simon \& Garfunkel's {\em The Sound of Silence}:
``{\it Hello darkness, my old friend~ I¡Çve come to talk with you
again}''. I hope that we will come to understand better our old dark friend
in due course.

%%%%%%%%%%%%%%%%%%%%%%%%%%%%%%%%%%%%%%%%%%%%%%%%%%%%%%%%%%%%%
\acknowledgments     %>>>> equivalent to \section*{ACKNOWLEDGMENTS}       
 
I thank Satoshi Miyazaki, Masahiro Takada, and Atsushi Taruya for
providing me their latest results that are reproduced in these
proceedings. I am also grateful to Ed Turner who generously gave me a
copy of Asimov's out-of-print book\cite{nightfall}.  Part of this work
is supported by JSPS (Japan Society for Promotion of Science)
Core-to-Core Program ``International Research Network for Dark Energy''
(DENET).

%%%%%%%%%%%%%%%%%%%%%%%%%%%%%%%%%%%%%%%%%%%%%%%%%%%%%%%%%%%%%
%%%%% References %%%%%

\bibliography{report}   %>>>> bibliography data in report.bib
\bibliographystyle{spiebib}   %>>>> makes bibtex use spiebib.bst

\end{document}